# ON THE APPLICATION OF THE GALLAVOTTI-COHEN FLUCTUATION RELATION TO THERMOSTATTED STEADY STATES NEAR EQUILIBRIUM

Denis J. Evans[a], Debra J. Searles[b] and Lamberto Rondoni[c]

[a]Research School of Chemistry, Australian National University, Canberra, ACT 0200, AUSTRALIA

[b]School of Science, Griffith University, Brisbane, Qld 4111, AUSTRALIA

[c]Dipartimento di Matematica and INFM, Politecnico di Torino, Torino, ITALY

## ABSTRACT

The fluctuation relation of the Gallavotti-Cohen Fluctuation Theorem (GCFT) concerns fluctuations in the phase space compression rate of dissipative, reversible dynamical systems. It has been proven for Anosov systems, but it is expected to apply more generally. This raises the question of which non-Anosov systems satisfy the fluctuation relation. We analyze time dependent fluctuations in the phase space compression rate of a class of N-particle systems that are at equilibrium or in near equilibrium steady states. This class does not include Anosov systems or isoenergetic systems, however, it includes most steady state systems considered in molecular dynamics simulations of realistic systems. We argue that the fluctuations of the phase space compression rate of these systems at or near equilibrium do not satisfy the fluctuation relation of the GCFT, although the discrepancies become somewhat smaller as the systems move further from equilibrium. In contrast, similar fluctuation relations for an appropriately defined dissipation function appear to hold both near and far from equilibrium.

# 1. INTRODUCTION

In 1993, Evans, Cohen and Morriss proposed a relation meant to describe the fluctuation properties of N-particle systems in nonequilibrium steady states that were maintained at constant energy by an appropriate deterministic time reversible *ergostat* [1]. This relation was based on heuristic theoretical arguments, and supported by computer simulation data. The authors of reference [1] borrowed an idea from the theory of nonlinear dynamical systems, that the expanding rates of trajectory separation along the unstable directions of the phase space in chaotic systems can be used to compute the steady state averages of smooth phase functions. For the first time they tested this idea in numerical calculations of nonequilibrium many particle systems (at that time, the same had been done using periodic orbit expansions, but only in calculations concerning low dimensional dynamical systems - see [2] for instance). Evans, Cohen and Morriss [1] used the symmetry properties of these expansion rates for time reversible systems, to propose a relation that we refer to as a steady state Fluctuation Relation (FR). Reference [1] motivated a number of papers in which various fluctuation theorems were derived or tested, the first of which were the Evans-Searles Transient Fluctuation Theorem (ESTFT) [3, 4], and the Gallavotti-Cohen Fluctuation Theorem (GCFT) [5] described in Sections 2 and 3.

A typical nonequilibrium system may consist of a relatively small number of particles that interact with each other and with an external field, $\mathbf{F}_e$, (the driven system). This system may be in thermal contact with a very much larger number of particles on which no external field acts. The reservoir particles could act as a heat bath effectively maintaining the smaller system of interest at a constant average temperature at least over the characteristic relaxation time required for the system of interest to relax to a (quasi) steady state. Although the whole system (driven system plus reservoirs) is Hamiltonian, the driven system by itself is non-autonomous and non-Hamiltonian.

One way of modeling such systems is to replace the large number of reservoir particles by a much smaller number of reservoir particles, each of which is subject to a time reversible deterministic force that imposes a constraint on their equations of motion. Among the most common constraints are those which constrain the internal energy of the system, called an "ergostat", and those that constrain the peculiar kinetic energy, called a "thermostat". These modified equations of motion were proposed simultaneously and independently by Hoover et al. [6] and Evans [7] in the mid 1980's and they have been studied theoretically and successfully employed in molecular dynamics computer simulations for two decades.

In the literature, the term 'thermostat' is sometimes used to refer to a constraint on the energy, kinetic energy or temperature of a system. In this paper we are careful to differentiate these, and only use the term 'thermostat' to refer to a constraint that is explicitly placed on the kinetic temperature of the system, and the term 'ergostat' to a constraint on the internal energy. If the constraints *fix* the kinetic temperature or the internal energy to a specified value at all times, so that these quantities do not fluctuate, we refer to the constraints as *isokinetic thermostats* or *isoenergetic ergostats*, respectively. Alternatively the constraints might allow the kinetic temperature or internal energy to fluctuate about a specified value, but ensure that there is no drift in that value. The Nosé-Hoover thermostat is an example of this type of thermostat. The way in which the constraint is incorporated into the equations of motion can also vary [8, 9]. If Gauss's principle of least constraint is satisfied by the constraint, then this is referred to as a Gaussian thermostat or ergostat.

Reference [1] considered a very long phase space (steady state) trajectory of a Gaussian ergostatted, (*i.e.* isoenergetic) N-particle system [10]. This long trajectory was divided into (non overlapping) segments of duration t. Along each of the trajectory segments, the instantaneous phase space compression rate, $\Lambda$

$$\Lambda \equiv \frac{\partial}{\partial \Gamma} \cdot \dot{\Gamma} \qquad (1)$$

was calculated. Here we denote the phase space vector describing the microstate (coordinates and momenta) of the N-particle system in d Cartesian dimensions by $\Gamma \equiv (\mathbf{q}_1, \mathbf{q}_2, ..\mathbf{q}_N, \mathbf{p}_1, ..\mathbf{p}_N)$. In [1], the dynamics is assumed to be chaotic and therefore the averaged value of the phase space compression rate computed along the trajectory segments of duration t, $\overline{\Lambda}_t$, can be considered to be a random variable whose probability distribution, $\text{Pr}(\overline{\Lambda}_t)$, can be constructed from the histogram of its observed values. Because of time reversibility of the dynamics, if the compression rate takes a value A, then it can also take the value -A, albeit with different probability. The FR tested in reference [1] states that:

$$\frac{1}{t} \ln \frac{\text{Pr}(\overline{\Lambda}_t = A)}{\text{Pr}(\overline{\Lambda}_t = -A)} = -A \qquad \text{for large t.} \qquad (2)$$

**Remark 1.** *One may find it odd to consider fluctuations in the phase space volume elements of mechanical systems. As a matter of fact, although the phase space compression rate is identically zero for Hamiltonian particle systems, it is non-zero for the (non-autonomous) dynamical systems obtained by restricting one's attention to an arbitrary subset of particles of that Hamiltonian system [11], (i.e. projecting out the coordinates and momenta of some of the particles). This is the case for the Hamiltonian system described above (driven system* *plus reservoirs), if the degrees of freedom of the reservoirs are projected out. One finds that heat is on average removed from the non-Hamiltonian reduced system, and that the corresponding phase space compression rate is nonzero and on average is negative [9].*

In reference [1], equation (2) was verified in nonequilibrium molecular dynamics computer simulations where a field, $\mathbf{F}_e$, induced a dissipative flux, $\mathbf{J}$. Because the system studied in [1] was maintained at constant energy using a Gaussian isoenergetic ergostat, $\Lambda(\Gamma) = [\beta \mathbf{J}](\Gamma) V \cdot \mathbf{F}_e$, and equation (2) can be written in an alternative but mathematically equivalent form,

$$\frac{1}{t} \ln \frac{\Pr(\overline{[\beta \mathbf{J}]}_t V \cdot \mathbf{F}_e = A)}{\Pr(\overline{[\beta \mathbf{J}]}_t V \cdot \mathbf{F}_e = -A)} = -A \qquad \text{for large t,} \qquad (3)$$

where for systems in d Cartesian dimensions,

$$[\beta \mathbf{J}](\Gamma) \equiv \frac{dN \mathbf{J}(\Gamma)}{2K(\Gamma)}. \qquad (4)$$

K is the (peculiar) kinetic energy. The dissipative flux $\mathbf{J}$ is defined in the usual way in terms of the adiabatic derivative of the internal energy, $H_0$, and the system volume V, [10],

$$\dot{H}_0^{ad}(\Gamma) \equiv -\mathbf{J}(\Gamma) V \cdot \mathbf{F}_e. \qquad (5)$$

This shows how for Gaussian isoenergetic dynamics, the instantaneous phase space compression rate can be equated with a physical quantity, which is recognizable as the (instantaneous) irreversible entropy production $\Sigma(\Gamma) = -[\beta \mathbf{J}](\Gamma) V \cdot \mathbf{F}_e = -\Lambda(\Gamma)$. This rate is a product of a thermodynamic force, $\mathbf{F}_e$, a thermodynamic flux, $\beta \mathbf{J}$ and the system volume V. In reference [1] both ways of writing the FR were exploited almost interchangeably.

In subsequent papers on fluctuation relations for nonequilibrium steady states, a range of different thermostatting methods have been considered, and in many of these $[\beta \mathbf{J}](\Gamma) V \cdot \mathbf{F}_e$ and $\Lambda(\Gamma)$ are not equivalent. For some steady state systems the long time *averages* $\overline{[\beta \mathbf{J}]}_t V \cdot \mathbf{F}_e$ and $\overline{\Lambda}_t$ are equal whereas the instantaneous values and finite time averages of $[\beta \mathbf{J}](\Gamma) V \cdot \mathbf{F}_e$ and $\Lambda(\Gamma)$ are not equal (e.g. Gaussian isokinetic thermostatted dynamics). This means that the probability ratios, $\dfrac{\Pr(\overline{[\beta \mathbf{J}]}_t V \cdot \mathbf{F}_e = A)}{\Pr(\overline{[\beta \mathbf{J}]}_t V \cdot \mathbf{F}_e = -A)}$ and $\dfrac{\Pr(\overline{\Lambda}_t = A)}{\Pr(\overline{\Lambda}_t = -A)}$, are not simply related, even asymptotically, and one cannot substitute one for the other in equations (2) and (3). In these cases there are at least two different fluctuation relations to consider: one for the phase space compression rate and the other for the dissipative flux, and the two relations might not be related. The phase space compression rate is the subject of equation (2) and of the FR of the GCFT [12] (see Section 2), while $[\beta \mathbf{J}](\Gamma) V \cdot \mathbf{F}_e$ is the subject of equation (3) and of the fluctuation relations of Evans and Searles for nonequilibrium steady states (see Section 3).

The FR inferred from (2) (in its dimensionless form [12]) has been obtained within the context of the GCFT [5], in which the average of the phase space compression rate is bounded by appropriate limits ($\overline{\Lambda}_t$ lies in the range $(-A^*, A^*)$, with $0 < A^* < \infty$ [13]), as discussed in Section 2. The GCFT has been proven for time reversible, dissipative, transitive Anosov systems, but it has been argued that the FR should apply more generally to systems of physical interest. Most of these systems can hardly be thought to be of the Anosov type in a mathematical sense, just as they cannot be considered ergodic. Therefore, the Chaotic Hypothesis (CH) was proposed in [5] in the hope that the class of systems satisfying the FR would be significantly larger than the class of Anosov systems. In a similar way the Ergodic Hypothesis justifies the equality of the time averages and ensemble averages of macroscopic variables to classes of system that are not strictly speaking ergodic. This raises the question of which non-Anosov systems satisfy equation (2), and the CH.

To address this question, we analyze time dependent fluctuations in the phase space compression rate for a class of thermostatted (not ergostatted) systems of particles that are at equilibrium or in steady states close to equilibrium. The particles are assumed to interact via potentials that are normally used to realistically model atomic and molecular interactions in statistical mechanics and molecular modeling. The equilibrium dynamics for this class of system does not generate the uniform phase space density of the microcanonical ensemble but rather generates the smooth but nonuniform phase space density of the canonical or isokinetic ensemble. Therefore, although there is no long time *average* phase space contraction or expansion, the instantaneous phase space compression rate fluctuates at equilibrium as the trajectory moves through phase space.

We begin by observing that numerical data do not seem to satisfy the FR of equation (2) if the state of the systems under consideration is thermostatted and close to an equilibrium state. The discrepancies in the test of this FR seem to become smaller as the external field increases and the system moves further from equilibrium [14-17]. Alternatively one may interpret the numerical data as an indication that for these systems the convergence times of equation (2) are so long that the fluctuations, which become smaller as the averaging time grows, become unobservable before equation (2) can be verified. In this paper we consider both possibilities, and provide two theoretical arguments to explain the numerical results of [14-17].

We consider the possibility that the GCFT does not apply to our systems. We try to identify reasons why the FR given in equation (2), and the CH, might not apply by analyzing the proof of the FR of the GCFT, under the assumption that it can be extended to equilibrium systems. Since the Anosov property is strictly violated even in systems in which equation (2) has been verified (e.g. [1, 18, 19]), we consider the characteristics of the Anosov property that our system does not have, but which are attributes of systems for which equation (2) is verified. We also consider the possibility that the FR of equation (2) is valid for our systems, but that it can only be verified numerically, at exceedingly long times. We find that this produces a difficulty in the derivation of the Green-Kubo relations. We show that for the correct Green-Kubo relations to hold, it is necessary for the fluctuations in the time-averaged phase space compression rate to converge to those of $[\beta \mathbf{J}](\Gamma) V \cdot \mathbf{F}_e$ at a sufficiently rapid rate.

In this paper we also note that the fluctuation relation of equation (3) has been numerically and experimentally verified for the class of systems considered here, at equilibrium and far from equilibrium. Furthermore the correct Green-Kubo relations can be derived from equation (3). We discuss how equation (3) can be obtained without recourse to the Anosov property (c.f. Section 3 and [20]). Combining these results, and in accord with the discussions on the meaning of the CH given in references [5, 21], we are led to the (perhaps surprising) conclusion that the CH does not apply to thermostatted systems.

In Section 2 we give a brief description of the CH and the GCFT, including a discussion of the conditions necessary for the GCFT. In Section 3 we describe the Evans-Searles FTs and highlight the differences between these theorems and the GCFT. In Section 4 we investigate the possibility of extending the proof of the GCFT to equilibrium dynamics, and discuss which violations of the Anosov property may differentiate our systems from those in which the FR of equation (2) has been verified. In Section 5 we show that the CH does not appear to be appropriate for a class of thermostatted systems that are in the linear response regime close to equilibrium. We show that, for those systems, equation (2) is in contradiction with the known Green-Kubo relations for transport coefficients in thermostatted systems. Section 6 summarizes our results.

## 2. THE GALLAVOTTI-COHEN FLUCTUATION THEOREM

In 1995 Gallavotti and Cohen [5] derived an equation equivalent to (2) within the framework of modern dynamical systems theory. For a dynamical system in phase space, $\mathcal{C}$, whose time evolution is governed by a map, S, they assumed the following (p. 936 of [5]):

(A) *Dissipation:* The phase space volume undergoes a contraction at a rate, on the average, equal to $D\langle\sigma(x)\rangle_+$, where 2D is the phase space $\mathcal{C}$ dimension and $\sigma(x)$ is a model-dependent "rate" per degree of freedom. (Note: For almost every initial condition, $\lim_{t\to\infty}\overline{\Lambda}_t$ in our notation equates to $-D\langle\sigma(x)\rangle_+$ in the notation of [5]. In other places in this paper we follow standard practice and use $\langle...\rangle$ to denote an ensemble average.)

(B) *Reversibility:* There is an isometry, i.e., a metric preserving map i in phase space, which is a map $i: x \to ix$ such that if $t \to x(t)$ is a solution, then $i(x(-t))$ is also a solution and furthermore $i^2$ is the identity.

(C) *Chaoticity:* The above chaotic hypothesis holds and we can treat the system ($\mathcal{C}$,S) as a transitive Anosov system.

The chaotic hypothesis that they proposed states (p. 935 of [5]):

**Chaotic Hypothesis (CH)**: *A reversible many-particle system in a stationary state can be regarded as a transitive Anosov system for the purpose of computing the macroscopic properties of the system*.

Gallavotti and Cohen then showed the following (p. 963 of [5]):

*"**Fluctuation Theorem:** Let (C, S) satisfy the properties (A)-(C) (dissipativity, reversibility, and chaoticity). Then the probability $\pi_\tau(p)$ that the total entropy production $D\tau t_0 \sigma_\tau(x)$ over a time interval $t = \tau t_0$ (with $t_0$ equal to the average time between timing events) has a value $Dt\langle\sigma(x)\rangle_+ p$ satisfies the large-deviation relation*

$$\frac{\pi_\tau(p)}{\pi_\tau(-p)} = e^{Dt\langle\sigma\rangle_+ p} \tag{6}$$

*with an error in the argument of the exponential which can be estimated to be $p, \tau$ independent.*

*This means that if one plots the logarithm of the left-hand side of (6) as a function of p, one observes a straight line with more and more precision as $\tau$ becomes large…"*

Note: if $\overline{\Lambda}_t = A$ then $p = -A/(D\langle\sigma\rangle_+)$.

The above theorem is known as the Gallavotti Cohen Fluctuation Theorem or GCFT for short [22]. It should be noted that the GCFT only refers to the phase space compression rate (called "entropy production" rate in [5], cf. p.936) and only to steady states. Apparently there is no direct requirement that the system should be maintained at constant energy, constant kinetic energy or even that it be maintained at constant volume. The GCFT only seems to require dynamics that is time reversible, smooth and to some degree hyperbolic, which makes the system behave as though it was a time reversible Anosov diffeomorphism. Therefore, equation (6), or equivalently its logarithm equation (2) should in principle apply to a rather wide class of dynamical systems, including, for instance, isothermal-isobaric as well as isoenergetic-isochoric N-particle systems, and also non-particle systems as long as their dynamics is sufficiently similar to that of reversible, transitive Anosov systems. As a matter of fact, Gallavotti and Cohen, on p. 939 of [5] state: "*The details of the models described here **will not** be used in the following, since our main point is the generality of the derivation of a fluctuation formula from the chaotic hypothesis and its (ensuing) model independence.*" They then give various examples of models for which the CH is expected to hold.

In a separate paper [13] Gallavotti pointed out that p should belong to an interval $(-p^*, p^*)$, where $p^*$ is the dynamically determined positive number, given below equation (2.7) in [13]. This important restriction on the application of the Theorem was not mentioned in [5]. In our present paper we include the statement of these bounds as a formal part of the GCFT. Thus the FR (6) of the GCFT is equivalent to equation (2), as long as one takes $\overline{\Lambda}_t = A = -D\langle\sigma\rangle_+ p$, with p in $(-p^*, p^*)$ and provided the system is dissipative (i.e. $\langle\sigma\rangle_+$ is positive).

**Remark 2.** *Equation (2) hides three fundamental aspects of the GCFT: a) it is only expected to be valid with A in a given interval $(-A^*, A^*)$, thus the domain of validity of the GCFT does not necessarily contain the full range of possible values of the fluctuations in time averages of the phase space contraction rate; b) if $A^*$ becomes zero the FR inferred from the GCFT is trivial; c) if convergence to the long time asymptotic expression (2) is too slow, verification of (2) would be impossible. In the latter two cases, the predictions of the GCFT may be formally correct, but inapplicable in practice (as discussed in Section 5). However, once this is clear, it is convenient to consider equation (2) as the prediction of the GCFT, and* *this is commonly done in the literature (e.g. [23]). The values of $A^*$ and the convergence rates to eq.(2) are normally difficult, if not impossible, to predict (cf. [13]) and will not be the subject of this paper.*

Reference [5] motivated tests (e.g. [24-26]) in different types of dynamical systems, where equation (2) or similar relations, were verified. The Gallavotti and Cohen work also motivated attempts at experimental verifications of the GCFT (see, for example, references [27, 28]), even though these experimental systems cannot be considered isoenergetic and the precise relationship between the instantaneous phase space compression rate and the measured properties in these experiments was not then known.

Quite obviously, realistic models of physical systems can hardly be expected to be transitive Anosov dynamical systems. Nevertheless, just as the mathematical notion of ergodicity is known to be violated by most common physical models and yet turns out to be extremely useful *for practical purposes,* the CH of [5] should be interpreted as saying that deviations from the transitive Anosov property cannot be observed at the macroscopic level. The CH then allows the use of the techniques of differentiable dynamics in the description of the steady states for a class of systems of physical interest, as long as one is only interested in the behavior of macroscopic observables. In particular, the CH allows one to describe the steady state of a given N-particle system as if it was given by a Sinai-Ruelle-Bowen (SRB) measure, *i.e.* a probability distribution which is smooth along the unstable directions of the dynamics, and which can be approximated by means of dynamical weights attributed to the cells of finer and finer Markov partitions.

However, at the present time the only test that has been attempted to determine whether a dissipative system satisfies the CH, is the numerical or experimental check of whether the system satisfies equation (2) (or an equivalent relationship) within accessible times. Different tests of the CH need to be designed to determine which physical systems it can be applied to. Until recently, all numerical evidence suggested that time-reversible steady state systems that were 'chaotic' to some degree, satisfy the CH. Indeed, strict chaos (meaning the presence of at least one positive Lyapunov exponent) did not even seem to be necessary for expressions such as equation (2) to be verified in numerical simulations of simple N-particle systems as long as the dynamics are sufficiently random [18, 19].

In this paper, we argue that the results of references [14-16] suggest that this view might not be correct. Reference [15] gives numerical evidence that thermostatted systems satisfy equation (2) at very high shear rates, while at small shear rates [14, 16] it becomes very problematic, or even impossible to verify it. As a matter of fact, the numerical results of [14, 16] suggest that as the system departs further from equilibrium, the data become more consistent with equation (2). If equation (2) (or an equivalent relation) affords the only possible test of the CH, these results appear in contradiction with the expectation that the CH should be satisfied better as the system approaches the equilibrium state and therefore becomes more chaotic (*i.e.* has a larger sum of positive Lyapunov exponents).

This is rather puzzling because there seems to be no obvious reason why thermostatted (constrained temperature) systems should behave so differently from ergostatted systems [29, 30]. Although there are many differences between the two, it is unclear what effect these could have on the applicability of the CH. Furthermore, it is our impression that the distance from equilibrium, or the precise amount of dissipation, which is invoked in the proof of the GCFT does not make any difference to the derivation of equation (2), as long as this dissipation is not exceedingly high [24]. Therefore, close to equilibrium and far from equilibrium thermostatted systems should not behave as differently as they do.

Thus, the domain of applicability of equation (2) and the CH is an open and quite intriguing question. In this paper we argue that equation (2) and the CH do not apply to thermostatted systems [31] that are near equilibrium. Also note that for hard discs or spheres, fixing the kinetic energy or the total energy are equivalent and therefore equation (2) is expected to apply to hard N-particle systems under these forms of thermostat since for hard systems both thermostats are in fact identical. Reference [24] gives evidence for the validity of the GCFT for one such system, i.e. for a system of thermostatted/ergostatted hard discs. However, if the kinetic energy is constrained using a Nosé-Hoover thermostat, our arguments imply that the CH does not apply to systems of hard core particles.

## 3. EVANS-SEARLES FLUCTUATION THEOREMS

A number of authors, inspired by [1, 5], have obtained a range of fluctuation relations for steady state systems which are similar in form to equation (2) but have different content and are applicable to either deterministic or stochastic systems. See, for example, references [32-36]. Still other authors, refined the GCFT, cf. references [37-39].

Independently of this activity, in 1994 Evans and Searles derived the first of a set of fluctuation theorems (ESFTs) for nonequilibrium N-particle systems which focused on a quantity $\Omega$, called the "dissipation function", rather than on the phase space compression rate, $\Lambda$ [3, 4]. For thermostatted or ergostatted nonequilibrium steady state systems the time average "dissipation function" is identical to the average rate of entropy absorption (positive or negative) by the thermostat. For homogeneously thermostatted systems the average entropy absorbed by the thermostat is equal and opposite to the so-called spontaneous entropy production rate defined in linear irreversible thermodynamics, $\Sigma = \sigma V$ where $\sigma$ is the "entropy source strength" defined in de Groot and Mazur [40]. Further for homogeneously thermostatted systems Evans and Rondoni [11] have recently shown that the entropy production rate is also equal and opposite to the rate of change of the fine grained Gibbs entropy. These ESFTs apply at all times to given ensembles [41] of transient trajectories (ESTFTs), or given ensembles of steady state trajectories in the long time limit [4] (ESSFTs). The form of the resulting FRs is similar to equation (2), but they contain different information since they are based on the statistics of the given ensembles of trajectories.

Jarzynski and Crooks have taken an approach similar to that of Evans and Searles, to calculate the free energy difference between equilibrium states [42, 43].

To derive the ESTFT one considers an ensemble of trajectories that originate from a known initial distribution (which may be an equilibrium or nonequilibrium distribution, it does not matter) and proceeds under the possible application of external fields and/or thermostats. One then obtains general transient fluctuation theorems (ESTFTs) stating that

$$\ln \frac{\Pr(\overline{\Omega}_t = A)}{\Pr(\overline{\Omega}_t = -A)} = At \qquad (7)$$

which is of similar form to (2) but where the time averaged phase space compression rate is replaced by the so-called time averaged dissipation function, $\overline{\Omega}_t(\Gamma)$, and Pr represents the probability which is influenced by the ensemble. In all the ESTFTs the time averages are computed from t=0 when the system is characterized by its initial distribution, $f(\Gamma,0)$, to some arbitrary later time t. The dissipation function depends on the initial probability distributions (different ensembles) and on the dynamics, and is defined by the equation,

$$\int_0^t ds\, \Omega(\Gamma(s);\Gamma(0)) \equiv \ln\left(\frac{f(\Gamma(0),0)}{f(\Gamma(t),0)}\right) - \int_0^t \Lambda(\Gamma(s))ds$$
$$= \overline{\Omega}_t(\Gamma(0))t \qquad (8)$$

for all positive times t.

For ergostatted dynamics conducted over an ensemble of trajectories which is initially microcanonical, the dissipation function is identical to the phase space compression rate,

$$\Omega(t) = -\Lambda(t) = -[\beta \mathbf{J}](t)V \cdot \mathbf{F}_e, \quad \text{when } dH_0/dt = 0, \tag{9}$$

while for thermostatted dynamics (both isokinetic and Nosé-Hoover), the dissipation function is subtly different,

$$\begin{aligned}\Omega(t) &= -\beta \mathbf{J}(t)V \cdot \mathbf{F}_e, \quad \text{constant T} \\ &\neq -\Lambda(t) = -\beta \mathbf{J}(t)V \cdot \mathbf{F}_e - \beta \dot{H}_0(t)\end{aligned}. \tag{10}$$

For isokinetic and isoenergetic dynamics, $\beta = 2K(\Gamma)/dN$ where d is the Cartesian dimension of the space in which the system exists. For Nosé-Hoover dynamics $\beta = 1/k_B T$ where $k_B$ is Boltzmann's constant and T is the absolute temperature appearing in the Nosé-Hoover equations of motion – see equation (20) below. It is clear that for constant temperature dynamics the dissipation function is different from the phase space compression rate. However, in all cases the magnitude of the time averaged dissipation function is equal (with probability one) to the magnitude of the average phase space compression rate since for thermostatted systems, $\lim_{t \to \infty}\left[\overline{\Omega}_t + \overline{\Lambda}_t\right] = \lim_{t \to \infty}\beta[\overline{\dot{H}_0}]_t = O(t^{-1})$. Thus $\lim_{t \to \infty}\overline{\Omega}_t = -\lim_{t \to \infty}\overline{\Lambda}_t = -\beta \lim_{t \to \infty}\overline{\mathbf{J}}_t V \cdot \mathbf{F}_e = \lim_{t \to \infty}\overline{\Sigma}_t$, where $\Sigma$ is the extensive entropy production that one would identify for near equilibrium systems from the theory of irreversible thermodynamics. The spontaneous entropy production is a product of the thermodynamic force $\mathbf{F}_e$ and the time average of its conjugate thermodynamic flux, $\beta \overline{\mathbf{J}}_t$ [40].

ESTFTs have been derived for an exceedingly wide variety of ensembles, dynamics and processes [4], and using both Liouville weights and Lyapunov weights [4, 44]. For example

4clean proseESTFTs have been derived for dissipative isothermal isobaric systems and for relaxation systems where there is no applied external field but where the system is not at equilibrium by virtue of its initial distribution $f(\Gamma,0)$. In all cases the ESTFTs have been verified in numerical experiments. Two ESTFTs have recently been confirmed in laboratory experiments: one involving the transient motion of a colloid particle in a moving optical trap[45]; another involving the relaxation of a particle in an optical trap whose trapping constant is suddenly changed [46]. One should not be surprised by the diversity of FTs - they refer to *fluctuations* and fluctuations are well known to be ensemble and dynamics dependent - even at equilibrium.

The ESTFT can be stated as follows:

**Theorem (Evans-Searles)**: *For any time reversible N-particle system, and for all positive times $t \in \mathbb{R}$, there exists a dissipation function $\overline{\Omega}_t$ and a smooth probability distribution $d\mu(\Gamma) = f(\Gamma)d\Gamma$ in phase space, such that:*

$$\frac{1}{t}\ln\frac{\Pr(\overline{\Omega}_t \in (A-dA, A+dA))}{\Pr(\overline{\Omega}_t \in (-A-dA, -A+dA))} = A + O(dA) \qquad (11)$$

*where* $\Pr(\overline{\Omega}_t \in (A-dA, A+dA))$ *is the probability assigned by* $\mu$ *to the set of initial conditions* $\Gamma$ *for which the dissipation* $\overline{\Omega}_t$ *lies in the range* $A \pm dA$.

It is interesting to observe that the probability measure $\mu$, *i.e.* its density, is not necessarily unique, and that different probability measures lead to the same result as long as $\ln \frac{f(\Gamma)}{f(S^t\Gamma)}$ , where $S^t$ is the time evolution operator, exists for all initial conditions $\Gamma$ in the support of $\mu$, and for all $t \in [0, \infty)$.

In contradistinction to the GCFT, these ESTFTs are not *only* true asymptotically in time but rather are valid for all times t.

Evans and Searles have also argued [4] that for transitive **chaotic** systems, where the **steady state exists and is unique**, the statistics of properties averaged over trajectory segments selected from a single steady state trajectory are equivalent to a carefully constructed ensemble of steady state trajectory segments [4, Section 2.2].

Assuming the arguments of [4] hold, one can derive asymptotic steady state FTs (ESSFTs) that apply to segments along a single trajectory, from the relevant ESFTs. The corresponding fluctuation formula for an ergostatted steady state system is then identical to (2), and contains the same information [21, 47]. That is, the FR of the GCFT and the ESSFT are the same for isoenergetic ergostatted steady state systems. It should also be noted that for systems that are not isoenergetically ergostatted, the predictions of the ESSFTs (given by equation (7) which becomes equivalent to (3)) are different in general from the corresponding predictions of the FR of the GCFT. This is because in general the dissipation function is different from the phase space compression rate. To check the validity of the ESSFTs, numerical simulations have been performed for various ensembles and dynamics, showing that numerical results are indistinguishable when sampling either from a single long steady state trajectory or from an ensemble of steady state trajectory segments [14].

**Remark 3.** *This equivalence of statistics requires a sufficiently long relaxation time to allow an accurate representation of the steady state, and long trajectory segments. Thus as is the case for the GCFT, the ESSFTs, in contradistinction to the ESTFTs, apply to steady states and are only valid at large t.*

The dissipation function that appears in the ESSFT for a single steady state trajectory is defined by (8), where the initial distribution function is the equilibrium distribution function generated by the same dynamics that is responsible for the steady state except that the dissipative field is set to zero [4]. This requires that the zero field system is ergodic and is at equilibrium.

# 4. EQUILIBRIUM FLUCTUATIONS

Equilibrium systems that exchange energy with their surroundings (such as those described by the canonical ensemble or the grand canonical ensemble) have fluctuations in their instantaneous energy and, their phase space distribution function is non-uniform in phase space (in contrast to that of the microcanonical ensemble). If the dynamics of such equilibrium systems are modeled by autonomous differential equations that contain terms that aim to mimic the energy exchange with the environment, the dynamics will not be Hamiltonian, and the phase space volumes will not be preserved. Therefore, the phase space compression rate of such systems can be non-zero at instants in time, although it will vanish on average. The models used in molecular dynamics simulations of such systems make use of thermostatting mechanisms which generally produce non-Hamiltonian dynamics, and generate equilibrium distribution functions $f(\Gamma)$ that are not uniform in phase space. For example, when applied to field-free Newtonian equations of motion, the Gaussian isokinetic thermostat generates the isokinetic distribution function, and the Nosé-Hoover thermostat generates the extended canonical distribution function [10]. These dynamics are non-dissipative, have an ensemble averaged phase space contraction which is zero, generate ensemble averaged state variables that are constant, and are invariant under a time reversal map (and therefore their properties will be time reversal invariant [10]). Yet since they are non-Hamiltonian and their phase space density is non-uniform, their instantaneous energy and phase space compression rates both fluctuate in time.

As a result of the time reversal invariance of all properties of the equilibrium state, we know that,

$$\frac{\Pr(\overline{\Lambda}_t(F_e = 0) = A)}{\Pr(\overline{\Lambda}_t(F_e = 0) = -A)} = 1 \quad \forall t. \tag{12}$$

This equation states that for all averaging times, the distribution of time averaged values of the phase space compression is precisely symmetric about zero. This is a special property of any equilibrium state. Equation (12) is a necessary but not sufficient condition for thermodynamic equilibrium - see references [48] for detailed discussions of equation (12) and how it is satisfied by both Gaussian isokinetic and Nose-Hoover thermostats.

Comparing equation (2) with equation (12), one can see that in equilibrium systems for which values of $A \neq 0$ are allowed at any finite averaging time t (no matter how large), equation (2) incorrectly predicts an asymmetry in the equilibrium distribution of time averaged values of the phase space compression rate. This would mean that in such systems, if they exist, some assumptions that are invoked in the derivation of (2) (i.e. in the derivation of the FR of the GCFT) must not hold (i.e. the CH of [5] does not apply). Below we consider the possibility that the systems modeled by equilibrium thermostatted dynamics are of this type.

For simplicity, let us focus on systems whose equations of motion are:

$$\begin{aligned} \dot{\mathbf{q}}_i &= \frac{\mathbf{p}_i}{m} \\ \dot{\mathbf{p}}_i &= \mathbf{F}_i - \alpha \mathbf{p}_i \end{aligned} \tag{13}$$

where $\alpha$ is a reversible thermostat multiplier that constrains the kinetic energy. Let us adapt the usual derivation of the FR to the case of non-dissipative systems of the kind (13), assuming that the CH holds for them [49]. In particular, let us consider the proof of the GCFT given by Ruelle in Section 3 of reference [37]. In the notation of [37], the dimensionless phase space compression rate at x over time $\tau$, $\varepsilon_\tau(x)$ is defined by Ruelle as,

$$\varepsilon_\tau(x) = \frac{1}{\tau e_f} \sum_{k=0}^{\tau-1} \log J(f^k x)^{-1} \tag{14}$$

where $e_f$ is the average phase space contraction per unit time, $f^k$ gives the time evolution of x, and J is the Jacobian of f with respect to the chosen metric. Comparing with the notation introduced above, we have $\tau \equiv t$, $\varepsilon_\tau(x) \equiv \overline{\Lambda}_t / \overline{\overline{\Lambda}}$ and $e_f \equiv -\overline{\overline{\Lambda}} \equiv \langle \Lambda \rangle$. Equation (14) excludes the cases with $e_f = 0$, and normalizes the phase space contraction rate so that $\varepsilon_\tau$ has a mean of 1. Nevertheless, the division by $e_f$ does not seem to be necessary for the proof in [37] to be carried out, and the calculations presented in Sections 3.6-3.9 of reference [37] can apparently be repeated even when the phase space contraction rate is not normalized. Assuming that this is the case, dynamics with $e_f = 0$ can be considered under the assumption that the CH holds for them, and Ruelle's derivation may then be repeated for the non-normalized phase space compression rate,

$$\varepsilon_\tau^o(x) = \frac{1}{\tau} \sum_{k=0}^{\tau-1} \log J(f^k x)^{-1} \tag{15}$$

instead of the *dimensionless* phase space compression rate $\varepsilon_\tau$. In general, $\varepsilon_\tau^o(x)$ takes a range of values for any system, even for equilibrium systems, but not for isoenergetic equilibrium systems which yield $\frac{1}{\tau} \sum_{k=0}^{\tau-1} \log J(f^k x)^{-1} = 0$ for any $\tau$ and any x. The range of admissible values of $\varepsilon_\tau^o(x)$ can be written as $[-p_o^*, p_o^*]$ which is symmetric about 0 due to time-reversibility. If our assumption is correct then, following the same steps of Ruelle's proof, one would obtain a relation formally identical to that reported in Section 3.9 of [37]. The only difference to Ruelle's result would be that this procedure does not yield a dimensionless expression, but whether $e_f$ is equal to zero or not would seem to make no difference to the adapted derivation. One could then write,

$$p_o - \delta \leq \lim_{\tau \to \infty} \frac{1}{\tau} \log \frac{\rho_f(\{x : \varepsilon_\tau^o(x) \in (p_o - \delta, p_o + \delta)\})}{\rho_f(\{x : \varepsilon_\tau^o(x) \in (-p_o - \delta, -p_o + \delta)\})} \leq p_o + \delta. \qquad (16)$$

Here, as in Ruelle [37], $\rho_f$ would be the probability, under the dynamics specified by f, that $\varepsilon_\tau^o(x)$ took on a value $p_o \in [-p_o^*, p_o^*]$, while $\delta > 0$ would be an arbitrarily small constant.

To obtain equation (16) the dynamics is assumed to be of the Anosov type, which implies that the phase space compression rate is a bounded function (Hölder continuous in [37]) and that $p_o^* < \infty$. If, however, $p_o^* \neq 0$ and an equilibrium system is considered, equation (16) is absurd, proving that these systems substantially violate the hypothesis on which the GCFT is based. Then the question as to which hypothesis is violated needs to be addressed.

Before focusing on this question, we note that the equilibrium systems considered in this paper are clearly not Anosov, however equation (2) (or (16)) has been tested numerically for a wide range of systems, none of which, to the best of our knowledge, meets all the conditions that the proof [37] requires. For instance, the models of [1, 15, 25, 26] are not expected to be uniformly hyperbolic; those of [18, 19, 24] have singularities; and the flat billiards of [18, 19] are not even chaotic (that is, have no positive Lyapunov exponents). But for them the FR has been shown to hold, and the CH considered appropriate to describe them. In other words, although the Anosov property is violated for these systems, this violation did not appear substantial. Therefore we must find possible reasons for the substantial violations of the Anosov property which would make the CH inapplicable to our systems (e.g. [14, 16, 17]).

For equilibrium Gaussian isokinetic dynamics, the value of $p_o^*$ ($\equiv A^*$), which delimits the range of admissible fluctuations, can be easily estimated. In fact, from equation (10), one finds that the average phase space contraction rate, $\overline{\Lambda}_t$, over a time t, is proportional to $(\Phi(t) - \Phi(0))/t$, where $\Phi(t)$ is the value of the interaction potential energy at time t, along the given phase space trajectory (see, e.g., refs.[50, 51]).

The Anosov condition implies that the instantaneous phase space compression rate is bounded, hence for Anosov equilibrium Gaussian isokinetic dynamics $\Phi$ must be bounded, and the asymptotic range of admissible fluctuations shrinks to zero. In this case the FRs for the phase space contraction rate, equations (2), (16), both make a completely trivial but correct prediction $\lim_{t \to \infty} \frac{1}{t} \ln \left[ \Pr(\overline{\Lambda}_t = 0) / \Pr(\overline{\Lambda}_t = 0) \right] = 0$! This prediction is "completely trivial" because provided $\Pr(\overline{\Lambda}_t = 0)$ is defined, the prediction is always true regardless the form of the probability distribution, or the CH, or indeed even time reversibility itself. The phase space compression rate is also bounded for non-Anosov isokinetic Gaussian thermostatted dynamics if the interaction potential is bounded. Strictly speaking, these systems are not Anosov, but they verify the FR of the GCFT, because that relation admits only $A = 0$. These systems may therefore look sufficiently similar to Anosov systems to be called "Anosov-like", and the CH may adequately characterize them.

However, if $\Phi$ is not bounded and the dynamics is isokinetic, as is commonly the case in NEMD models, the range of admissible fluctuations might not shrink to the unique zero value, it might be finite, or even infinitely large. In such cases, the FR for the phase space contraction rate is incorrect.

For Nosé-Hoover thermostatted dynamics, which is a much better model of a real thermostatted system, the range of possible values for the phase space compression factor is always infinite, regardless of whether the potential function is bounded or not. Hence, in this case the FR for the phase space contraction is also either incorrect or trivial.

Moreover, if the possible violation of CH is attributed to the singularities of the phase space contraction rate, this violation persists at small fields, where the new nonequilibrium phenomena cannot remove the effect of the singularities. In fact, for sufficiently small fields, the probability distribution for averages of the phase space contraction rate is expected to be little different from that at equilibrium. This could explain the results of [14, 16, 17], in which the FR of the GCFT could not be verified.

Many other scenarios are consistent with the available numerical evidence. For instance, one subtle, but dramatic, violation of the CH could be inferred from the fact that the number of positive finite-time Lyapunov exponents fluctuates along phase space trajectories of the thermostatted systems. This indicates that the continuous splitting of the tangent space of our dynamics, required by the Anosov condition, does not hold even approximately for our systems. If this is the case, close to equilibrium systems would violate the CH for the same reason.

Another possible scenario concerns the times for convergence of equation (2). If these times are too long the CH will be invalid in a practical sense, as discussed in Section 5.

The discrepancy between (2) and (12) for isothermal systems can be contrasted to the agreement between equations (3), (7) and (12), for an ensemble of isothermal systems. Applying the ESTFT for arbitrary phase functions (equation (4.19) of [4]) to the dissipative flux, J, gives,

$$\frac{1}{t}\ln\frac{\Pr(\bar{J}_t = A)}{\Pr(\bar{J}_t = -A)} = -\beta A V F_e, \qquad (17)$$

where the trajectory segments begin from the isokinetic equilibrium ensemble and proceed for a time t, under zero field $F_e = 0$, isokinetic thermostatted dynamics. However since the external field is zero, equation (17) predicts that at equilibrium time averages of the dissipative flux are as expected, equally likely to be positive or negative, regardless of the duration of the averaging time.

In summary, the FR given in equation (2) does not apply to *thermostatted* equilibrium systems, while (3) and (7) do. However for *isoenergetic* equilibrium states, equation (2), the ESTFT (equation (7)), ESSFT, equation (17) and equation (3) all make correct statements about the equilibrium symmetry of fluctuations. For equation (2) this is due to the fact that A can only take the value 0 in this system, while the ESTFT yields

$$\frac{1}{t}\ln\frac{\Pr(\overline{[\beta J]}_t = A)}{\Pr(\overline{[\beta J]}_t = -A)} = -AVF_e. \tag{18}$$

In (18) J refers to the component of **J** that is parallel to $\mathbf{F}_e$, so when the field is zero, the ESTFT states that time averages of the thermodynamic flux [βJ] are equally likely to be positive as negative, regardless of the averaging time. This is obviously a correct statement.

In the next section we discuss the application of the FR to thermostatted near equilibrium steady states.

## 5. THE APPROACH TO EQUILIBRIUM

Consider a thermostatted or ergostatted dissipative system described by the equations of motion,

$$\dot{\mathbf{q}}_i = \frac{\mathbf{p}_i}{m} + \mathbf{C}_i \cdot \mathbf{F}_e$$
$$\dot{\mathbf{p}}_i = \mathbf{F}_i + \mathbf{D}_i \cdot \mathbf{F}_e - \alpha \mathbf{p}_i \qquad (19)$$

For typical interatomic forces $\mathbf{F}_i$, the system is time reversible and chaotic. Gallavotti (in 1996) [52] was the first to point out that, (at least in the case of ergostatted dynamics - see below), the GCFT (and hence equivalently the ESSFT), can be used to derive the well known Green-Kubo relations for linear (near equilibrium) transport coefficients [53]. Later Searles and Evans [54] showed that the ESSFT for thermostatted systems could also be used to derive correct Green-Kubo relations for linear transport coefficients [55]. We now argue that in the Nosé-Hoover thermostatted case where,

$$\dot{\alpha} = \frac{1}{Q}(2K - dNk_BT) = \frac{2K_0}{Q}(K/K_0 - 1) \equiv (K/K_0 - 1)/\tau^2 \qquad (20)$$

(where $Q = 2K_0\tau^2$ is related to the arbitrary relaxation time $\tau$, of the thermostat, K is the peculiar kinetic energy and $K_0$ is some chosen fixed value of the peculiar kinetic energy), the FR for phase space compression (equation (2)) is not applicable, since it is inconsistent with the correct Green-Kubo relations for linear transport coefficients.

The Nosé-Hoover extended canonical (equilibrium) distribution is:

$$f_c(\Gamma,\alpha) = \frac{\exp(-\beta(H_0 + \tfrac{1}{2}Q\alpha^2))}{\int d\alpha \int d\Gamma \exp(-\beta(H_0 + \tfrac{1}{2}Q\alpha^2))}, \quad (21)$$

from which the distribution of $\{\alpha\}$ can be obtained by integration,

$$f_c(\alpha) = \frac{\exp(-\tfrac{1}{2}\beta Q\alpha^2)}{\int d\alpha \exp(-\tfrac{1}{2}\beta Q\alpha^2)} = \sqrt{\frac{\beta Q}{2\pi}} \exp(-\tfrac{1}{2}\beta Q\alpha^2) \quad (22)$$

which is Gaussian with a variance $\sigma_\alpha^2 = 1/(\beta Q)$ [56]. Assuming equations (21) and (22) hold for our equilibrium systems, the distribution of $\overline{\alpha}_t$ is also Gaussian because it is just the time average of $\alpha$.

The variance of $\overline{\alpha}_t$ for systems that are in near-equilibrium steady states can also be considered. For simplicity we assume $\mathbf{F_e} = (F_e, 0, 0)$, however the following can be readily adapted to apply more generally. From the equations of motion (19, 20) we see that the rate of change of the extended Nosé-Hoover Hamiltonian $H_0' \equiv H_0 + \tfrac{1}{2}Q\alpha^2$, is

$$dH_0'/dt = -JVF_e - dNk_BT\alpha. \quad (23)$$

The external field contributes to the fluctuations in the phase space compression rate. This contribution cannot be expected to be Gaussian except when long time averages are made, near the mean of the distribution.

32From (23) we see that,

$$(H_0'(t) - H_0'(0))/t \equiv \Delta H_0'(t)/t = -\bar{J}_t V F_e - dN k_B T \bar{\alpha}_t. \qquad (24)$$

So the variance of time averages of $\alpha$ contains, to leading order, two contributions,

$$\sigma^2_{\bar{\alpha}_t} = (\sigma^2_{\Delta H_0'(t)}/t^2 + V^2 F_e^2 \sigma^2_{\bar{J}_t})/(dN k_B T)^2. \qquad (25)$$

Here, and below, we used the fact that many properties, including $\sigma^2_{\bar{\alpha}_t}$, must be even functions of the field. Because we assume a steady state, in the long time limit, $\sigma^2_{\Delta H_0'(t)}$ is independent of t. In fact near equilibrium $\lim_{t \to \infty} \sigma^2_{\Delta H_0'(t)} = 2k_B T^2 C_V' + O(F_e^2)$ where $C_V'$ is the extensive (O(N)), isochoric specific heat of the extended system [10].

From [54] we know that for sufficiently long times,

$$t \sigma^2_{\bar{J}_t} = 2L(F_e) k_B T / V + O(F_e^2 t^{-1} N^{-1}) \qquad (26)$$

where $L(F_e)$ is the zero-frequency Green-Kubo transform of the dissipative flux, $L(F_e) = \beta V \int_0^\infty dt \left\langle \left(J(0) - \langle J \rangle_{F_e}\right)\left(J(t) - \langle J \rangle_{F_e}\right) \right\rangle_{F_e}$. We also know from the Green-Kubo relations that, $\lim_{F_e \to 0} L(F_e) = L(0)$ is the linear transport coefficient defined by the linear constitutive relation,

33$$\lim_{F_e \to 0} \lim_{t \to \infty} \frac{-\bar{J}_t}{F_e} = L(0). \tag{27}$$

We note that at nonzero fields $L(F_e)$ has no simple relation to the nonlinear transport coefficient for the process [54].

Substituting (26) into (25) and using the relationship between $\sigma^2_{\Delta H'_0(t)}$ and $C'_V$ gives, at long times and small fields,

$$\sigma^2_{\bar{\alpha}_t} = 2k_B T^2 C'_V / (dN k_B T t)^2 + 2V F_e^2 L(F_e) / (t k_B T (dN)^2) + O(F_e^4 t^{-2} N^{-1})$$
$$= O(t^{-2} N^{-1}) + O(F_e^2 t^{-1} N^{-1}) \tag{28}$$

In the weak field limit the mean of $\alpha$ is,

$$\overline{\overline{\alpha(F_e)}} = -\beta \bar{\bar{J}} V F_e / (dN) \sim \beta L(F_e = 0) V F_e^2 / (dN) = O(F_e^2) \quad \text{for small } F_e. \tag{29}$$

Now we would like to consider the limit $t \to \infty$, so that we can simultaneously:

- ensure that t is as large as required by the FR of equation (2);
- ensure that the Central Limit Theorem (CLT) applies, and hence near the mean, the distribution of $\bar{\alpha}_t$ can be described by a Gaussian;
- generate fully converged Green-Kubo integrals.

However as we increase the integration time t, the variance of the distribution of $\bar{\alpha}_t$ gets ever smaller. This implies that for fixed $F_e$, the mean of the distribution of $\bar{\alpha}_t$, which has a fixed mean value $\bar{\bar{\alpha}}$, moves more and more standard deviations away from zero. This means that symmetric fluctuations, like $\pm\bar{\bar{\alpha}}$, which are the object of equation (2), need not be described by a Gaussian distribution at long times with fixed $F_e$. To ensure that the typical fluctuations of $\bar{\alpha}_t$, namely $\pm\bar{\bar{\alpha}}$, have their distribution described accurately by a Gaussian, we propose to take the following limits simultaneously: $t \to \infty$ AND $F_e \to 0$ while keeping $\bar{\bar{\alpha}}/\sigma_{\bar{\alpha}_t} = r$ constant.

Substituting from equations (28) and (29) gives,

$$r^2 = \frac{\bar{\bar{\alpha}}^2}{\sigma_{\bar{\alpha}_t}^2} = \frac{F_e^4}{\frac{a}{t^2} + \frac{bF_e^2}{t}} \qquad (30)$$

where a,b are constants independent of t, $F_e$. Solving this quadratic equation for $F_e^2$ shows that we must take the limit $t \xrightarrow[(F_e = ct^{-1/2})]{} \infty$, where c is a constant. To simplify notation we denote this limit simply as $\lim\limits_{\substack{t \to \infty \\ F_e \to 0}}$.

**Remark 4.** *A more conservative procedure which ensures that the distribution is Gaussian for typical fluctuations, is to keep* $\bar{\bar{\alpha}}/\sigma_{\bar{\alpha}_t} = rt^{(1-m)}$ *constant where* $m > 1$ *[57]. This implies that the limit is taken such that* $F_e = \left(ct^{-1/2}\right)^m$.

Using this procedure, the distribution of $\alpha$ will be Gaussian near the typical values of A required by equation (2) for longer and longer t. This gives:

$$\lim_{\substack{t\to\infty \\ F_e \to 0}} \frac{1}{t} \ln \frac{\Pr(\overline{\alpha}_t = A)}{\Pr(\overline{\alpha}_t = -A)} \sim \frac{2A\overline{\overline{\alpha(F_e)}}}{\sigma^2_{\overline{\alpha}_t} t}$$

$$= \frac{dNA}{(TC'_V/(VF_e^2 L(0)t) + 1)} \qquad (31)$$

$$= \frac{dNA}{(Tnc'_V/(c^2 L(0))+1)}$$

In this equation $c'_V$ is the intensive specific heat per particle and $n=N/V$ is the number density of the system.

We now show that (31) contradicts the result inferred from equation (2). If we assume (2) is correct both at and near equilibrium, then, for some sufficiently large time $\tau_\Delta(F_e)$, the difference between the two sides of (2) is $\Delta$ or smaller than $\Delta$. In the weak field regime, close to equilibrium, one may expect that $\tau_\Delta(F_e) = \tau_\Delta(F_e = 0) + O(F_e^2)$. If this is the case, equation (2) implies,

$$\lim_{\substack{t\to\infty \\ F_e \to 0}} \frac{1}{t} \ln \frac{\Pr(\overline{\alpha}_t = A)}{\Pr(\overline{\alpha}_t = -A)} = dNA . \qquad (32)$$

This result is in contradiction with (31). Since in our limit $\lim_{\substack{t\to\infty \\ F_e \to 0}}$, the Central Limit Theorem and the Green-Kubo relations cannot be called into question, we conclude that (32) is incorrect and therefore the FR inferred from the (2) cannot be applied to Nosé-Hoover thermostatted systems. Similar conclusions are reached even if, for any fixed $\Delta > 0$, $\tau_\Delta(F_e)$ cannot be simply expressed as $\tau_\Delta(F_e = 0) + O(F_e^2)$ but rather does not grow faster than $O(1/F_e^2)$.

The other possibility suggested by references [14, 16, 17] is that there is a $\Delta > 0$ such that $\tau_\Delta(F_e) \to \infty$ faster than $O(1/F_e^2)$ as $F_e \to 0$. In this case, although (2) might be formally correct, it is not able to be verified at low enough fields, and, above all, it cannot be used to derive the GK relations. In fact the validity of the CLT is required for GK to be derived (see, e.g., references [17, 52]), but the CLT does not apply to the time-dependent probability distribution functions if the times grow faster than $O(1/F_e^2)$, as discussed above.

We now repeat these arguments assuming that the ESSFT holds. The steady state version of the ESFT for thermostatted systems states,

$$\frac{1}{t}\ln\frac{\Pr(\overline{J}_t = A)}{\Pr(\overline{J}_t = -A)} = -\beta A V F_e \qquad \text{for large t.} \qquad (33)$$

Using the same procedure as above, we can apply the CLT to show that the distribution of $\overline{J}_t$ will be Gaussian near the mean and at typical values of the fluctuations, at long t. That is, taking the long time, small field limit as above, i.e. keeping $\left|\overline{\overline{J}}\right|/\sigma_{\overline{J}_t} = r$, and applying the CLT to the dissipative flux one obtains

$$\lim_{\substack{t \to \infty \\ F_e \to 0}} \frac{1}{t} \ln \frac{\Pr(\bar{J}_t = A)}{\Pr(\bar{J}_t = -A)} \sim \frac{2A\bar{\bar{J}}}{\sigma_{\bar{J}_t}^2 t}. \tag{34}$$

Note that taking the limits simultaneously, so that we keep $\left|\bar{\bar{J}}\right|/\sigma_{\bar{J}_t} = r$ constant implies $F_e^2 t$ is constant. This is the same limit as that taken in equation (31). Combining equation (33) and (34), and using the linear constitutive relation for the linear transport coefficient (equation (31)) gives,

$$L(F_e = 0) \equiv \lim_{\substack{t \to \infty \\ F_e \to 0}} \frac{-\bar{\bar{J}}}{F_e} = \lim_{\substack{t \to \infty \\ F_e \to 0}} \frac{\beta V t \sigma_{\bar{J}_t}^2}{2}. \tag{35}$$

After some tedious manipulations of the integrals (see [54]) we find that

$$L(F_e = 0) = \beta V \int_0^\infty ds \left\langle J_\gamma(0) J_\gamma(s) \right\rangle_{F_e = 0}, \quad \gamma = x, y, z. \tag{36}$$

The notation $\left\langle .. \right\rangle_{F_e = 0}$ denotes an ensemble average taken over thermostatted trajectories with the external field set to zero. This is the correct Green-Kubo expression for a linear transport coefficient $L(F_e = 0)$, of a thermostatted system [10], and shows that a FR for the dissipative flux does not suffer from the difficulties of the FR for the phase space compression rate.

# 6. CONCLUSION

Our theoretical analysis indicates that the Fluctuation Relation inferred from the Gallavotti Cohen Fluctuation Theorem does not apply to thermostatted equilibrium states, or near-equilibrium thermostatted steady states. The same holds for the isothermal isobaric systems, and for any other steady state system whose energy is not a strict constant of the motion. For systems at or near equilibrium, equation (2) only gives correct, useful predictions for the time averaged fluctuations in the phase space compression factor if the energy is fixed.

This conclusion is supported by the inability of computer simulation calculations to verify equation (2) for non constant energy, near equilibrium particle systems [14, 16, 17], in contrast to the ease with which the FR for the dissipation functions are verified in these systems. The calculations show that the discrepancies between the data and the predictions of equation (2) become greater in relative magnitude as the dissipative field strength is reduced and the steady state approaches the equilibrium state. No such difficulties are encountered in the tests of the ESFTs.

Because (2) is an asymptotic relation expected to be valid only at sufficiently long times, one could argue that the computer data have not been tested at sufficiently long times. However in tests of (2) and the integrated version of (2) [14], the computer tests have been carried out at times which are very long indeed – of the order of 1000 Maxwell relaxation times! Even at this long time the disagreement for equation (2) is an order of magnitude larger than the numerical errors (see Figure 3b of reference [14]). However the data cannot rule out the possibility that at some extremely long time which is completely inaccessible to computer simulation or experiment, the two sides of (2) do indeed converge to the same value.

Our theoretical analysis does not provide definitive reasons as to why the FR inferred from the GCFT (2) cannot be applied to the thermostatted steady state systems and equilibrium systems we study. We have discussed a number of possibilities:

- the errors in (2) may go to zero more slowly than the standard deviation of the fluctuations in the time-average of the phase space compression rate, in which case the fluctuations may become unobservable before the FR (2) is verified;

- the range of fluctuations in phase space compression within which the GCFT is valid, may be zero at equilibrium; and/or

- the Chaotic Hypothesis may be substantially violated by any system which is not maintained at fixed energy.

In relation to the first possibility: away from equilibrium the standard deviation of the probability distributions appearing in equation (2) is of order $t^{-1/2}$. If the difference between both sides of equation (2) vanishes more slowly than $t^{-1/2}$, then as the time increases the fluctuations become unobservable before equation (2) can be verified. In such a case it would be impossible (even in principle) to confirm the validity of (2).

In contrast, all numerical and experimental tests have validated the ESFTs within accessible observation times. Moreover, when a corresponding theoretical analysis is made of the near equilibrium fluctuations, this analysis yields the well known Green-Kubo expressions for the relevant linear transport coefficients. This indicates that for thermostatted nonequilibrium steady states, a FR for the dissipative flux (like equation (3)) is useful, in contrast to the FR in terms of the phase space contraction given by equation (2). The practical relevance and utility of the ESTFTs and ESSFTs has recently been confirmed in laboratory experiments [45, 46, 58].

Recently van Zon and Cohen have shown that the phase function that is the subject of their "Generalized Fluctuation Theorem" [23] fails to satisfy a relationship of the form given in (2). Evans has recently pointed out [59], that this property corresponds to the phase space contraction considered in the GCFR. The conclusion of van Zon and Cohen is therefore quite consistent with the present paper.

We find it hard to understand why changing the constraint mechanism from a Gaussian ergostat to a thermostat, can have such drastic effects, since for *ergostatted* systems, the GCFT seems to correctly describe equilibrium and near equilibrium fluctuations. This puzzle is not resolved by comparing the Lyapunov spectra for thermostatted and ergostatted systems. At the same thermodynamic state point, the two spectra are remarkably similar.

We interpret our results as implying that the natural measures of thermostatted systems at, or close to equilibrium, are quite different from the SRB measures, from which the GCFT is derived. This is undoubtedly related to the fact that at equilibrium, instantaneous phase space compression rates of the thermostatted dynamics can be non-zero, although the implications of this fact are not fully understood yet.

We can demonstrate this quite clearly through the following example. Consider a Gaussian isokinetic thermostatted system (rather than the Nosé-Hoover thermostatted systems considered previously in this paper). For such a system where the equations of motion take the form given in (19), consider the particular case where $\mathbf{C}_i = \mathbf{0}$. We can separate the contributions to the thermostat multiplier, $\alpha = \sum_i [(\mathbf{F}_i + \mathbf{D}_i \cdot \mathbf{F}_e) \cdot \mathbf{p}_i]/(2K_0)$, that are due to the external field from those that are intrinsic to the field free system [17]. In such a case one can show that if one rewrites equation (2) so that it refers only to the fluctuations in the phase space compression rate that are *explicitly* due to the external field, we then obtain the correct description of both the at and near equilibrium fluctuations.

Furthermore, in the above example, as the field is increased, the full Gaussian isokinetic thermostat multiplier $\alpha$, will be increasingly dominated by the second, field dependent term. In that case even if we do not separate the explicit field dependent contribution from the phase space compression rate it is clear that as the field increases the argument of the FR will be increasingly dominated by the explicitly field dependent term. Hence the relation given in equation (2) will be approximated more and more accurately by the FR of equation (3) as the field strength is increased (provided that negative fluctuations remain observable as the field increases). The fact that the error in (2) decreases as the field increases is not because the CH is more likely to apply at large fields (in fact the opposite is true) but is related to the simple fact that, at larger field strengths, fluctuations in the phase space compression rate more closely approximate those of the dissipation function, $\Omega$, which is the subject of the ESFTs. These fluctuations are well behaved and satisfy the ESFTs. This is consistent with the numerical results [14, 16, 17], and may explain the better numerical verification of equation (2) for some systems as the field strength increases, and chaoticity decreases [14-17]. All the arguments considered above for the thermostatted systems lead to the conclusion that equation (2), and the CH on which it is based, are not of practical use, even if (2) eventually converges to a correct result. In other workds, our analysis suggests that the CH is not an appropriate characterization of thermostatted systems, except perhaps for ergostatted isoenergetic systems.

# ACKNOWLEDGEMENTS

We would like to thank the Australian Research Council for their support of this project. LR gratefully acknowledges financial support from the Scientific Office of the Italian Embassy in Australia. We would also like to acknowledge helpful comments from David Ruelle, Rainer Klages, Federico Bonetto, Giovanni Gallavotti, G. Benettin and L. Galgani.